\documentclass[%
 reprint,
 superscriptaddress,
 amsmath,amssymb,
 aps,
 twocolumn,
prb,
]{revtex4-2}

\usepackage{graphicx}% Include figure files
\usepackage{dcolumn}% Align table columns on decimal point
\usepackage{bm}% bold math
\usepackage{multirow}
\usepackage{bbding}
\usepackage{color}
\newcommand{\beginsupplement}{%
        \setcounter{table}{0}
        \renewcommand{\thetable}{\arabic{table}}%
        \setcounter{figure}{0}
        \renewcommand{\thefigure}{S\arabic{figure}}%
     }

\newcommand{\Cr}{CrCl$_2$(pyz)$_2$ }
\newcommand{\Mo}{MoCl$_2$(pyz)$_2$ }
\begin{document}

%\preprint{APS/123-QED}

\title{Interaction Driven Topological Phase Transition in Monolayer CrCl$_2$(pyrazine)$_2$} % Force line breaks with \\

\author{Xuecong Ji}
\thanks{These authors contributed equally to this work.}
\affiliation{Beijing National Laboratory for Condensed Matter Physics, \\
and Institute of Physics, Chinese Academy of Sciences, Beijing 100190, China}
\affiliation{University of Chinese Academy of Sciences, Beijing 100049, China}

\author{Jiacheng Gao}
\thanks{These authors contributed equally to this work.}
\affiliation{Beijing National Laboratory for Condensed Matter Physics, \\
and Institute of Physics, Chinese Academy of Sciences, Beijing 100190, China}
\affiliation{University of Chinese Academy of Sciences, Beijing 100049, China}

\author{Changming Yue}
\affiliation{Department of Physics, University of Fribourg, 1700 Fribourg, Switzerland}

\author{Zhijun Wang}
\affiliation{Beijing National Laboratory for Condensed Matter Physics, \\
and Institute of Physics, Chinese Academy of Sciences, Beijing 100190, China}
\affiliation{University of Chinese Academy of Sciences, Beijing 100049, China}

\author{Hua Wu}
\affiliation{Laboratory for Computational Physical Sciences (MOE), State Key Laboratory of Surface Physics, and Department of Physics, Fudan University, Shanghai 200433,China}
\affiliation{Collaborative Innovation Center of Advanced Microstructures, Nanjing 210093,China}

\author{Xi Dai}
\email{daix@ust.hk}
\affiliation{Department of Physics, Hong Kong University of Science and Technology, Kowloon 999077, Hong Kong, China}

\author{Hongming Weng}
\email{hmweng@iphy.ac.cn}
\affiliation{Beijing National Laboratory for Condensed Matter Physics, \\
and Institute of Physics, Chinese Academy of Sciences, Beijing 100190, China}
\affiliation{University of Chinese Academy of Sciences, Beijing 100049, China}
\affiliation{Songshan Lake Materials Laboratory, Dongguan, Guangdong 523808, China}

\date{\today}% It is always \today, today,

\begin{abstract}
The quadratic band crossing points (QBCPs) at Fermi level in two-dimension have been proposed to be unstable under electron-electron interaction. The possible interaction driven states include quantum anomalous Hall (QAH) state and various nematic ordered states. In this work, motivated by the discovery of ferromagnetic van der Waals layered  metal-organic framework CrCl$_2$(pyrazine)$_2$, we theoretically propose that the single layer of CrCl$_2$(pyrazine)$_2$ might realize one or some of these interaction driven states based on the QBCP protected by $C_4$ symmetry. By introducing the short-range density-density type repulsion interactions into this system, we have found the phase diagram depending on different interaction range and strength. The exotic phases include the staggered chiral flux state manifesting QAH effect, the site-nematic insulator and the site-nematic Dirac semimetal state. The QAH state is robust against perturbations breaking the QBCP but it is weakened by increasing temperature. The metal-organic framework is tunable by changing the transition-metal elements, which might improve the gap size and stability of this interaction induced QAH state. 
\end{abstract}

\maketitle

%\tableofcontents

\section{\label{sec:level1}Introduction}
% In the quantum anomalous Hall (QAH) insulator~\cite{QAHE_haldane_PhysRevLett.61.2015,QAHE_qi_PhysRevB.74.085308}, the ground
% state breaks time-reversal symmetry but does not break the lattice translational symmetry.
% The ground state has a bulk insulating gap, but has chiral edge states.The QAHE have been realized in magnetically-doped or  intrinsic ferromagnetic topological  insulator  at extremely low temperature.
% In a 2D  interacting fermions system, the quadratic band crossing point(QBCP) is unstable under weak interaction, leading to the spontaneous symmetry breaking of rotation symmetry or time-reversal invariance ~\cite{sunPhysRevB.78.245122,WenxiaogangPhysRevB.77.235125}.
% recently  the layer material \Cr has been synthesized .The calculation and experimental result all confirm the ferromagnetic configuration.While the conductivity measurement suggest a insulated ground state conflicted with the DFT calculated metallic ground state~\cite{pedersen2018formation,wuhuaD1TC00518A,Xie_2019}.Based on first-principle calculation we find that there is a QBCP at $\Gamma$ point, we perform the variational Hartree-Fock calculations to analysis how the QBCP change with  repulsion interaction .

In the past decades, the quantum anomalous Hall (QAH) states have aroused great interests in the  condensed matter physics community~\cite{QAHE_haldane_PhysRevLett.61.2015,QAHE_qi_PhysRevB.74.085308,qahe_science_MTI,PhysRevLett.107.186806_HgCrSe,PhysRevB.84.075129_pyrochlore_iridates,PhysRevB.87.125405_gated_ilayer-graphen,qahe_science.1234414,doi:10.1080/00018732.2015.1068524}. Different from quantum Hall effect, the chiral edge state of QAH insulator is resulted from the breaking of time reversal symmetry (TRS) without introducing external magnetic field. Recently, the QAH effect has been realized in magnetically-doped~\cite{qahe_science_MTI,qahe_science.1234414,Checkelsky2014} or intrinsic two-dimensional ferromagnetic topological insulator~\cite{Gong_2019,doi:10.1126/sciadv.aaw5685,doi:10.1126/science.aax8156} at extremely low temperature. The improvement of critical temperature is highly needed for extensive study and potential application, but it is quite challenge. One of the reason is that in these realizations the bulk gap is opened and limited by the effective strength of spin-orbit coupling (SOC) of the  inverted valence and conduction bands, which leads to nonzero Chern number.
%QAHE一些  MnBiTe 的实验文章

%指出其他方案的不足 都要依赖与soc 铁磁性强弱 能带反转 限制了体能大小 ，这里的 由相互作用 导致 
 This band gap limitation might be avoided in the interaction induced QAH insulator state proposed in various schemes.~\cite{Topo_mott_insulator_PhysRevLett.100.156401,inter_qahe_kagome_PhysRevLett.117.096402,inter_QAHE_PhysRevB.91.125139,sunPhysRevB.78.245122,FENG20211384,sun2009topological,wu2016diagnosis,RG_qahe_inter_PhysRevB.89.201110,PhysRevB.98.125144,PhysRevB.96.205412} Especially, the quadratic band crossing point (QBCP) in two-dimensional systems is proved to be unstable against arbitrarily weak short-range repulsive interactions~\cite{sun2009topological,RG_qahe_inter_PhysRevB.89.201110,Tsai_2015,wu2016diagnosis,PhysRevB.98.125144,PhysRevB.96.205412,wzj2022}. The resulting gaped phases are QAH or nematic state depending on the interaction parameters and temperature. The QBCP semimetal state requires the protection of fourfold or sixfold rotational symmetry ~\cite{sun2009topological,wu2016diagnosis}. In the typical checkerboard lattice model satisfying $C_4$ rotation symmetry, a fairly large next nearest-neighbor (NNN) hopping parameter is assumed. Otherwise, the QAH phase would be dominated by the nematic phase even in low temperature, or the QAH order parameters are too small to  be observed ~\cite{sun2009topological,wu2016diagnosis}.

Recently, the ferromagnetic van der Waals layered metal-organic framework \Cr has been synthesized and the Curie temperature is found to be 55 K~\cite{pedersen2018formation}. The first-principles calculations~\cite{wuhuaD1TC00518A,Xie_2019} and experimental measurements~\cite{pedersen2018formation,doi:10.1126/science.abb3861} all confirm its ferromagnetic ground state. While the first-principles calculations give a semimetal band structure~\cite{wuhuaD1TC00518A,Xie_2019}, the conductivity measurement~\cite{pedersen2018formation} suggests an insulating ground state. In this work, we proposed \Cr to be the material realizing the two-dimensional spinless QBCP model on the checkerboard lattice. To achieve the insulating ground state, the short-range Coulomb interactions have been taken into consideration under mean-field level. In this realistic system, the amplitude of NNN hopping is fairly small compared to the nearest-neighbor (NN) hopping. To explore more possible phases in this system, we added two types of NNN interactions have been considered based on the crystal structure of \Cr. The QAH state appears in the phase diagram when the strength of NNN interaction is close to the NN hopping amplitude. Besides, we found another type of nematic phase with Dirac points. In addtion, we find the site-nematic Dirac semimetal state is also possible. 
%These different phases  proposed in this model could be achieved in  this metal-organic materials family.

\section{\label{sec:level2}First-principles calculations}
The first-principles calculations are based on density functional theory (DFT)~\cite{dft_PhysRev.140.A1133} using the Vienna Ab-initio Simulation Package (VASP)~\cite{vasp_PhysRevB.54.11169}. The wave function is  expressed with the plane-wave basis set and the exchange and correlation effect are described by the generalized gradient approximation (GGA) with the Perdew-Burke-Ernzerhof(PBE) functional~\cite{gga_pbe1_PhysRevLett.77.3865,gga_pbe2_PhysRevLett.78.1396}. The kinetic energy cutoff for the plane-wave basis is set to 500 eV. In the self-consistent calculation, the Monkhorst–Pack~\cite{MP_PhysRevB.13.5188} k-mesh of 9 × 9 × 1 are used for the  Brillouin zone integration. We use GGA+$U$ method~\cite{ggau_PhysRevB.57.1505} with Hubbard $U$ = 4.0 eV and Hund exchange $J$ = 0.9 eV for the 3$d$ orbitals of Cr~\cite{wuhuaD1TC00518A}.

\begin{figure}[!tb]
	\centering
	\includegraphics[width=0.45\textwidth]{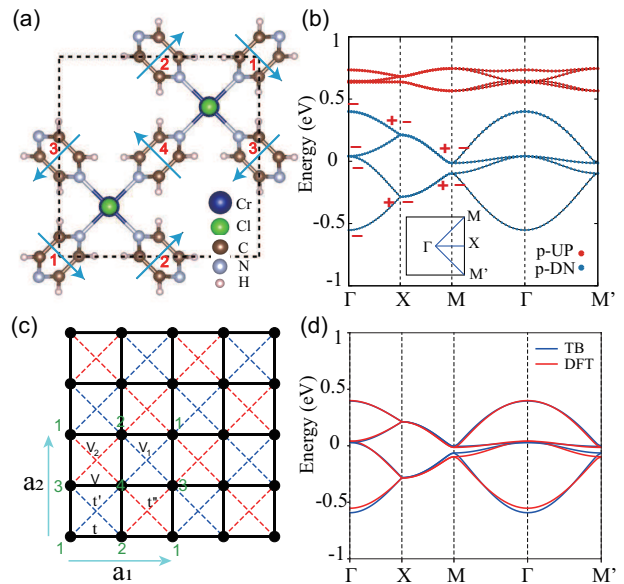}% Here is how to import EPS art
	\caption{(a) The top view of monolayer CrCl$_2$(pyz)$_2$. The arrows represent the in-plane projection of pyrazine rings' normal vectors. We label the four rings from 1 to 4. (b) The DFT band structure of ferromagnetic \Cr. The red and blue dots represent the spin up and down $p$-orbital weight from  C and N atoms. On each TRIM, the parities of the four states around Fermi level are labled with + and - for even and odd, respectively. The inset is the Brillouin zone(BZ) and high symmetry points. (c) The graphic plot of the low-energy effective model. The black solid lines represent hopping amplitude $t$ and interaction strength $V$ among NN sites. The blue dashed lines represent NNN hopping $t'$ and interaction $V_1$ with Cr atoms as intermediate site, while the red dashed lines represent $t''$ and $V_2$ without intermediating Cr. (d) The fitted band structure of the four bands near the Fermi level with effective tight-binding Hamiltonian. }
	\label{fig:dft}
	
\end{figure}

The monolayer \Cr is crystallized in space group P4/nbm\cite{Xie_2019} with the four-fold rotation axis perpendicular to the layer and pass through Cr atoms. There are four pyrazine rings in each primitive cell as shown in Fig.~\ref{fig:dft}(a). The arrows are the in-plane projections of the normal vectors of pyrazine rings. They are along the diagonal lines of the cell.
Within the GGA+$U$ first-principles calculation, the ferromagnetic order on Cr gives out a semimetal-type band structure, as shown in Fig.~\ref{fig:dft}(b). The four bands around the Fermi level are dominantly contributed by the spin-down $p$-orbitals from the C and N atoms on the pyrazine rings, separating from the spin-up conducting bands. The gap between spin-up bands and spin-down bands doesn't change with the on-site Hubbard-$U$ but the bandwidth of the spin-down bands around Fermi level is slightly decreased by increasing the value of  $U$. The SOC effect is fairly weak because these bands are coming from the $2p$ orbitals of C and N atoms and they are strongly spin-polarized. The polarization of pyrazine rings is opposite to Cr ions and it is also looked as a ferrimagnetic state~\cite{pedersen2018formation}.

The QBCP at $\Gamma$ point is protected by the $\Tilde{C}_4$ symmetry ($\Tilde{C}_4 = \{C_4|1/2,0,0\}$). A slight breaking of $\Tilde{C}_4$ but preserving $\Tilde{C}_2$ would split the QBCP into two Dirac points with linear dispersion\cite{wu2016diagnosis}. The parities of the Bloch states on the time-reversal invariant momenta (TRIM) around Fermi level are labeled in Fig~\ref{fig:dft}(b), from which we can deduce that a gaped state of the system without further band inversion at $M$ would lead to a QAH state. From the topological quantum chemistry analysis~\cite{TQC_Bradlyn2017,PhysRevB.97.035139,GAO2021107760}, these four bands come from the elementary band representation (EBR) $A_u@4f$. The Wyckoff position $4f$ exactly sits at the center of four pyrazine rings, which indicates the low energy bands come from the effective molecular orbitals of pyrazine rings. The site symmetry group of the ring center is $2/m$, in which the two fold rotational axis is along the diagonal axis of the primitive cell. Thus, the $A_u$ representation on each site is embed by any $p$ orbital perpendicular to this in-plane C$_2$ rotation axis. For such a reason, we choose the $p$ orbitals parallel to the arrows in Fig.~\ref{fig:dft}(a) to construct the effective model Hamiltonian.

\begin{figure*}[!htb]
	\includegraphics[width=1\textwidth]{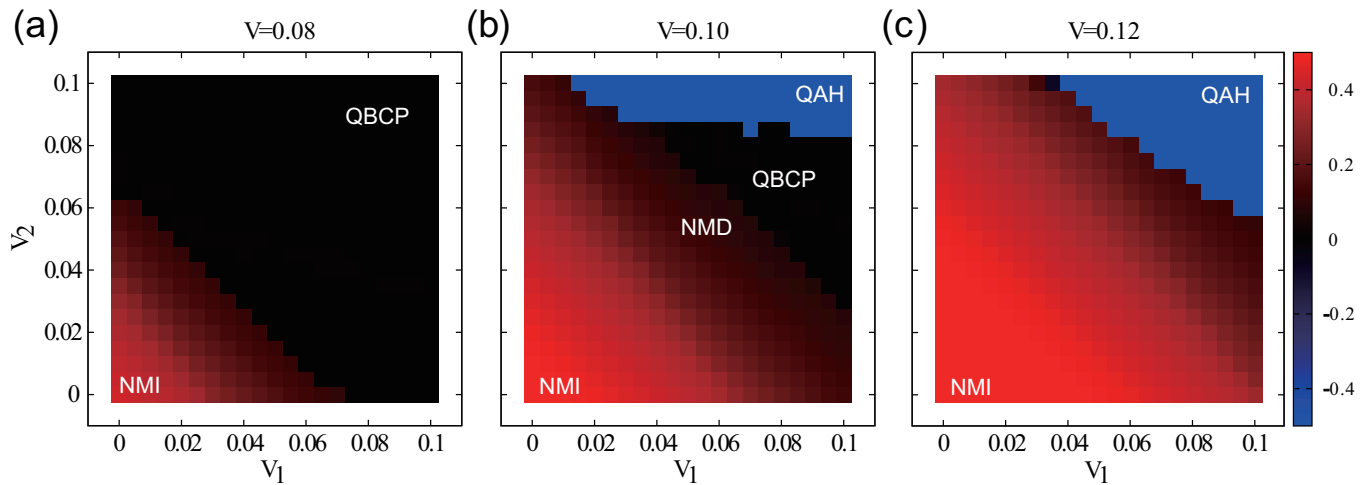}% Here is how to import EPS art
	\caption{ (a-c) The mean-field phase diagram under different NN interaction $V$ values. The order parameters of nematic phase, including site nematic insulator (NMI) and nematic Dirac semimetal (NMD), is colored in red and QAH order parameter is colored in blue. The ground states with negligible QAH and site-nematic order parameters are colored in black, where the QBCP at $\Gamma$ persists. }
	\label{fig:phase}
\end{figure*}

\section{\label{sec:level3} model and results }
We construct the four-band tight-binding (TB) model by fitting the band structure to the ab-initio results. The model parameters contain NN hopping amplitude $t$,  NNN hopping $t'$ and  $t''$ with and without intermediating Cr atoms, respectively, as shown in ~\ref{fig:dft}(c). This model is similar to the common checkerboard lattice model except that there are four sites in one primitive cell. The extension of primitive cell in \Cr is due to the four different norm vector directions of pyrazine rings. The directed vector on lattice sites leads to a non-symmorphic space group because all rotational axes, including the out-plane four-fold rotation axis and the in-plane two-fold rotation axis, pass through Cr atoms but the inversion centers are at the pyrazine ring centers. This non-symmorphic property doubles the primitive cell compared to the checkerboard lattice and causes the degeneracy along XM high-symmetry line. We construct the TB model respecting all symmetrical operations mentioned above. Besides, the four bands around Fermi level are fully spin polarized in spin down channel. By ignoring SOC terms, the model can be constructed without considering the spin degree of freedom. Thus, this spinless model preserves the time reversal symmetry as manifested by the real number hopping parameters. Compared to the two-site checkerboard lattice model, the QBCP on M point now folds to $\Gamma$ due to the doubling of the unit cell as shown in Fig.~\ref{fig:dft}(c). As we know the QBCP is unstable against short-range interaction in the checkerboard lattice model, in the following we add the interaction terms in this doubled model to check which phase is the ground state under various interaction parameters when QBCP is gapped or split. 

The interaction driven phase transition on the two-site checkerboard lattice has been discussed in Ref.~\cite{sun2009topological,wu2016diagnosis} and in those works only the NN repulsion $V$ is investigated. Similarly, if only $V$ is considered in this case, the realistic hopping parameters, namely $t'/t=-0.187,t''/t=-0.070$, will make the area of $V$-driven QAH phase quite small and a tiny increasing of temperature would kill it. Furthermore, the chemical potential does not pass the QBCP exactly, which weakens the instability of QBCP. To find a stable and relatively large portion of QAH phase region, two types of NNN repulsion interaction are considered to compete with the NN repulsion (Fig.~\ref{fig:dft}). Therefore, the total Hamiltonian reads as 
\begin{equation}
\begin{split} 
    \hat{H}_{\rm T} &= \hat{H}_{\rm TB} + V\sum_{\langle ij\rangle} \hat{n}_i \hat{n}_j \\
            &+ V_1 \sum_{\langle\langle ij\rangle\rangle'} \hat{n}_i \hat{n}_j + V_2\sum_{\langle\langle ij\rangle\rangle''} \hat{n}_i \hat{n}_j
\end{split}
\end{equation}
where $\hat{n}_i$ is the density operator on site $i$, $\langle ij\rangle$ stands for the pair of NN sites, and $\langle\langle ij\rangle\rangle'$ ($\langle\langle ij\rangle\rangle''$) stands for the NNN site-pair with (without) Cr atom at the intersite. Details about the model can be found in the Supplementary Materials.

This model was solved by using the Hartree-Fock variational method~\cite{PhysRevB.103.035427}. The mean-filed Hamiltonian with variational parameters on every bond is used to provide the single-particle ground state. By searching the minimal point of total energy in the variational parameter space, one can get the ground state within mean-field approximation level. 
\begin{figure}[!tb]
	\includegraphics[width=0.45\textwidth]{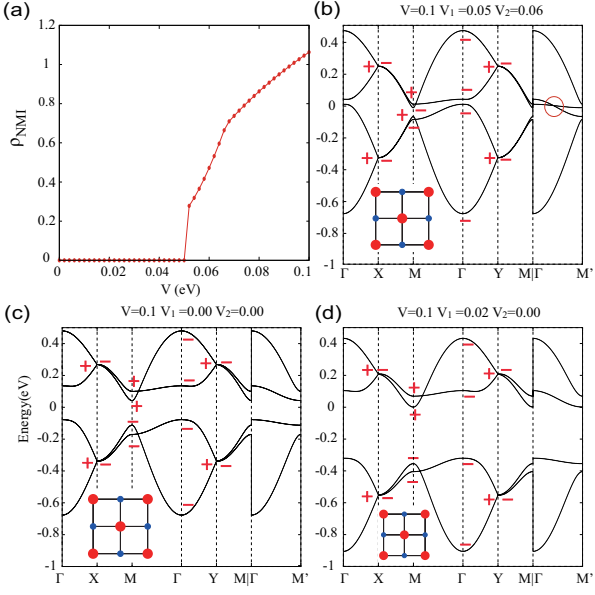}
	\caption{(a)The order parameter of NMI state evolves with $V$, and the other two interaction parameters are $V_1=V_2=0$. The phase transition occurs at $V=0.053$eV from the normal state to the NMI phase. Fixing $V$=0.1 eV, the dependence of the band structures on $V_1$ and $V_2$ is shown in (b) for NMD state, (c) and (d) for NMI state. The insets are graphical plots of the corresponding ordered state in real space with the dot size representing the charge disproportion among the sites. The red circle in (b) marks the Dirac point. On each TRIM, the even and odd parities of the occupied bands are labeled with + and -, respectively.
	 }
	\label{fig:NMIQAH}
\end{figure}
Fig.~\ref{fig:phase} and~\ref{fig:NMIQAH} show the phase diagram of the model within various interaction parameters. When the NNN repulsion interactions $V_1$ and $V_2$ are not applied, the observable phase transition occurs at $V=0.053$eV (Fig.~\ref{fig:NMIQAH}a) from the normal state to the site-nematic insulator (NMI) phase. The order parameter describing the nematic phase is defined as 
\begin{equation}
    \rho_{\rm NMI} = \Bar{n}_1 - \Bar{n}_2
\end{equation}
where $\Bar{n}_i$ is the charge density on site $i$. At this phase, the charge redistribution on the four sites in one unit cell breaks the four-fold rotation into two-fold one. Two sites connected by one of the diagonal line accommodate more electrons than the other two as schematically shown in the inset of Fig.~\ref{fig:NMIQAH}. The energy gap in that phase is quite large and the occupied states do not hold nontrivial topological property as can be seen from the parities of the occupied state in Fig.~\ref{fig:NMIQAH}(c). This topological trivial phase can be well understood by the EBR analysis. Without four-fold rotation, the two sites with more electron accumulation can form a complete EBR in the new phase, as being topologically equivalent to an atomic insulator. Note that the QAH phase is hard to be observed in this model with only NN repulsion because $t'/t$ is fairly small as shown in Fig.~\ref{fig:phase}(a), which is consistent with previous results.~\cite{sun2009topological,wu2016diagnosis}

Introducing the NNN repulsion leads to the competition between NMI and other states. As we can see from Fig.~\ref{fig:phase}, with a relatively small $V$ value, the growing of $V_1$ and $V_2$ recovers the normal state with QBCP within the mean-field calculation. This is easy to understand because the increase of charge on the diagonal two atoms would save energy when only NN repulsion is applied. But as the NNN repulsion increasing, it would cost more energy to maintain an imbalanced charge distribution.
When introducing the NNN repulsion, the NMI still exist in the area where the NNN repulsion is weak. We also find a nematic Dirac semimetal (NMD) state as a transition state between NMI and normal state as shown in Fig.~\ref{fig:phase}(b). In that phase, the redistribution of charge tends to cause a band inversion at $M$ point to recover an atomic insulator and the splitting of QBCP into two Dirac points along $\Gamma M'$ [Fig.~\ref{fig:NMIQAH}(b)]. As the parameters are approaching the NMI phase, the Dirac points move towards the $M'$ point. Finally they merge at $M'$ or $M$ and open a gap. The band inversion happens at the same time and make the NMI phase trivial. Notice that the NMD state has not been found in the two-site checkerboard lattice model because it needs the fine tuning of the interaction strength and hopping amplitude.

\begin{figure}[!tb]
	\includegraphics[width=0.5\textwidth]{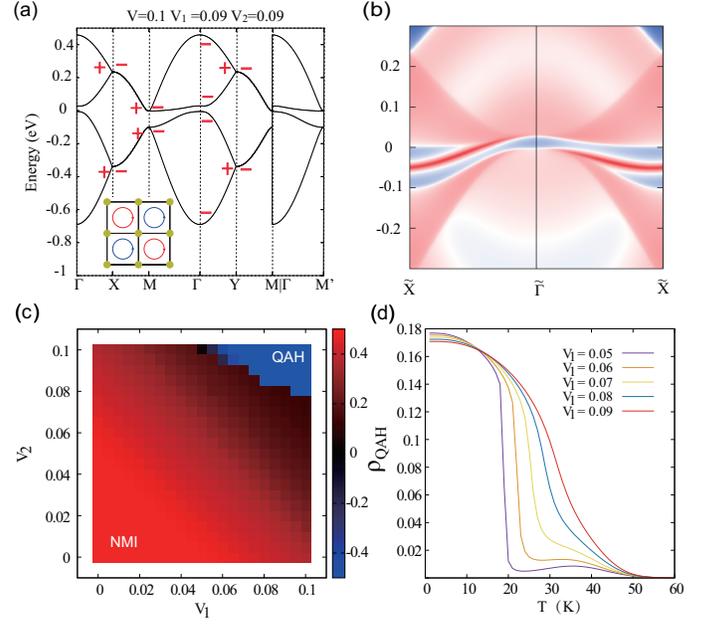}
	\caption{(a)The band structures of QAH phase, the charge is distributed equally, but there exist imaginary part of hopping on the nearest bonds, which forms a staggered flux pattern as shown in the inner window. (d) The chiral edge state of the QAH state in (a).(c) The phase diagram of the C$_4$ breaking model with $m=0.0025$. (d) The order parameter of QAH state evolve with temperature. The other two interaction parameters are $V=0.1$, $V_2=0.09$.}
	\label{fig:qah}
\end{figure}

Interestingly, for some proper $V$ values, the time reversal symmetry breaking QAH phase would appear spontaneously in a considerable large parameter space. The order parameter describing the QAH phase is defined as

\begin{equation}
    \rho_{\rm QAH} ={\rm Im}\sum_{\langle ij\rangle}\sum_{n,k} \left( \langle p_i|\psi_{n,k}\rangle \langle \psi_{n,k}|p_j\rangle  \right)
\end{equation}
%修改定义 
where $|p_{i}\rangle$ means the local orbital on site $i$ in the homecell and $|\psi_{n,k}\rangle$ is the eigen function of $n$th band at $k$. $\langle ij\rangle$ runs over NN bond pair of $\langle 12\rangle,\langle 24\rangle,\langle 43\rangle,\langle 31\rangle$. $n$ is the occupied bands and $k$ samples the whole BZ. The nonzero order parameter indicates the existence of imaginary part of NN hopping, which causes a staggered chiral flux pattern and breaks the time reversal symmetry. The flux configuration is shown in Fig~\ref{fig:NMIQAH}(a) with its band structure. In QAH phase, all in-plane $C_2$ symmetries are broken but $C_4$ and inversion are preserved. This set of symmetry operations do not protect any two-fold degeneracy at $\Gamma$ and the system becomes gaped. Due to the $C_4$ symmetry, charges still distribute equally on the four sites, and the occupied two bands can not form a complete EBR. The occupied states must possess non-trivial topological properties. As show in Fig.~\ref{fig:NMIQAH}(a), we label the parity on every TRIM points for the bands. The parity of Chern number can be calculated by using the following formula~\cite{formula_PhysRevB.98.081101}
\begin{equation}
    (-1)^{\rm Chern} = \prod_{\textbf{K}_{\rm TRIM}} \xi(\textbf{K}_{\rm TRIM})
\end{equation}
where $\xi(\textbf{K}_{\rm TRIM}) = \prod _n \xi_n(\textbf{K}_{\rm TRIM})$ and $\xi_{n}(\textbf{K}_{\rm TRIM})$ is the parity of the $n$th occupied Bloch state at TRIM point $\textbf{K}_{\rm TRIM}$. The chiral edge state~\cite{WU2018405} of the QAH phase is shown in Fig.~\ref{fig:NMIQAH}(b).

To illustrate the stability of this interaction induced QAH state, we further investigate the perturbations of breaking $C_4$ symmetry and increasing temperature. We find that although the QBCP is protected by the C$_4$ symmetry, the QAH phase can still emerge under proper interaction parameters when C$_4$ breaking terms are added to the model. Fig.~\ref{fig:qah}(c) shows the phase diagram when including the following perturbation term 
\begin{equation}
    \delta H(k) = m\sigma_z\otimes\sigma_z,
\end{equation}
which is $k$-independent as long as the perturbation is onsite term. Therefore, the ground state only preserves the inversion symmetry because the QAH order and nematic charge distribution appear simultaneously. In Fig.~\ref{fig:qah}(b), it is shown that the QAH phase is suppressed by increasing temperature. The model with larger interaction strength will lead to stabler QAH ground state against the temperature, which is consistent with previous study ~\cite{sun2009topological,wu2016diagnosis}.

\begin{table}[t]
	\centering
	\caption{\label{tab:table2}Symmetry properties for the various ordered states.}
		\begin{tabular}{c|c|c|c|c}
			\hline
			\hline
			& $I$ &$\Tilde{C}_{4,001}$ & $C_{2,110}$ & $\Tilde{C}_{2,010}$\\ 
            \hline
            QBCP&\Checkmark&\Checkmark&\Checkmark&\Checkmark\\
            \hline
	        QAH&\Checkmark&\Checkmark&\XSolidBrush&\XSolidBrush\\
	        \hline
	        NMI&\Checkmark&\XSolidBrush&\Checkmark&\XSolidBrush\\
			\hline
			NMD&\Checkmark&\XSolidBrush&\Checkmark&\XSolidBrush\\
			\hline
		\end{tabular}
		\label{table:1}
\end{table}

In Table~\ref{tab:table2}, we summarized the preserved and broken symmetries of all the phases mentioned above. There exists other types of phases such as stripe charge order when $V_1$ or $V_2$ is set to be much larger than $V$. But under a reasonable assumption that the interaction with the shorter distance would have the larger strength, we do not discuss these cases in this work.

\section{\label{sec:level4} Conclusion}
In summary, the first-principles calculations of the ferromagnetic van der Waals layered  metal-organic framework \Cr exhibit a quadratic band crossing at $\Gamma$ point when $C_{4}$ is preserving. To describe the low energy bands, we established a $C_4$ symmetric TB model by using the effective $p$-type molecular orbitals sitting on the center of each pyrazine ring. Due to the insulating ground state indicated by the experiments, we added the short-ranged repulsive interactions on the TB model. There are fruitful ordered phases to be the ground state as we tune the interaction parameters, including QAH, NMI and NMD states. The QAH state is prominent by the spontaneously appearing of time-reversal breaking staggered chiral flux. The QAH phase is stable against the C$_4$ breaking perturbation in the TB model. These ordered phases may appear in this family of metal-organic frameworks with different transition-metal elements or bond length, which can tune several things including the interaction strength or hopping amplitude.

\begin{acknowledgments}
This work was supported by the National Natural Science Foundation of China (Grant No. 11974395, 11925408, 11921004 and 12188101), the Ministry of Science and Technology of China (Grant No. 2018YFA0305700), the Chinese Academy of Sciences (Grant No. XDB33000000) and the Informatization Plan of Chinese Academy of Sciences (Grant No. CAS-WX2021SF-0102), the Hong Kong Research Grants Council (Project No. GRF16300918 and No. 16309020) and the Center for Materials Genome.

\end{acknowledgments}

\bibliographystyle{apsrev4-2.bst}
\bibliography{main.bib}% Produces the bibliography via BibTeX.

\clearpage 
\begin{widetext}
\beginsupplement{}
\setcounter{section}{0}
\renewcommand{\thesubsection}{\arabic{subsection}}
\renewcommand{\thesubsubsection}{\alph{subsubsection}}

\section{Tight-binding model}
The details of TB model in this paper are as follows
\begin{equation}
\begin{aligned}
   H_{\rm TB} = \begin{pmatrix}
        m&{\gamma}_1&{\gamma}_2&{\gamma}_3\\
        {\gamma}^{\ast}_1&m&{\gamma}_4&{\gamma}_2\\
         {\gamma}^{\ast}_2& {\gamma}^{\ast}_4&m& {\gamma}_1\\
          {\gamma}^{\ast}_3& {\gamma}^{\ast}_2& {\gamma}^{\ast}_1&m\\
        \end{pmatrix}\\
    {\gamma}_1 = t(1+e^{-ik_x})\\
    {\gamma}_2 = t(1+e^{-ik_y})\\
    {\gamma}_3 = t'(1+e^{-ik_x-ik_y})+t''(e^{-ik_x}+e^{-ik_y})\\
    {\gamma}_4 = t'(1+e^{ik_x-ik_y})+t''(e^{ik_x}+e^{-ik_y})
\end{aligned}
\end{equation}
where $m$  is the  on site energy, $t$ is the NN hopping amplitude, $t'$ and $t''$ are the NNN hopping amplitude passing through Cr atoms or not. The values of them fitted with DFT band structure are shown in Table ~\ref{tab:s1}

\begin{table}[!htb]
	\centering
	\caption{\label{tab:table1}The hopping parameters in the TB model.}
		\begin{tabular}{cccc}
			\hline
			\hline
			$m$&$t$&$t'$&$t''$\\
			\hline
			-0.03462&0.12359&-0.02313&-0.0086\\
			\hline
		\end{tabular}
		\label{tab:s1}
\end{table}

\section{Hartree-Fock method}
After mean  field  approximation, the the four-fermion operators in $\hat{H}_{\rm T}$ can be wirte as  single particle operator with parameter as equation \ref{mean}. So apply the Hartree-Fock variational method, we can transform the the  $\hat{H}_{\rm T}$ to the following mean field Hamiltonian with variational parameters $\lambda_a$ and   single particle operators $\hat{O}_a$ has been used to determine the variational ground state wave function 
\begin{equation}
\begin{aligned}
H_{\rm MF} & = H_{\rm TB} + \sum_a \lambda_a \hat{O}_a\\
&=H_{\rm TB}+ \sum_{l, \boldsymbol{k}} h_{l}^{\prime} \hat{O}_{2 l}(\boldsymbol{k})+h_{l}^{\prime \prime} \hat{O}_{2 l+1}(\boldsymbol{k})+\sum_{m, \boldsymbol{k}} \epsilon_{m} \hat{n}_{m}(\boldsymbol{k})
    \end{aligned}
\end{equation}
 and 
 \begin{equation}
 \begin{aligned}
\hat{O}_{2l}(k)&=\hat{c}^{\dagger}_j(k)\hat{c}_k(k)e^{-ik\cdot\tau}+h.c.\\
\hat{O}_{2l+1}(k)&=i(\hat{c}^{\dagger}_j(k)\hat{c}_k(k)e^{-ik\cdot\tau}-h.c.)
 \end{aligned}
\end{equation}
   The index $m,l$ runs over all sites and bond.  where the site $i,j$ are connected by the bond $l$,$\tau$ is the connected vector.The operators $\hat{O}_a$ is chosen this form to make $\hat{O}_a$ to be Hermitian and $\lambda_a$ to be real.
 To be more specific, there are 4 sites and 16 bonds in one primitive cell in the model of \Cr.  In this case, there are 36 variational parameters to be determined.
  The  mean field Hamiltonian has been used to determine the variational ground state wave function. The ground state $|\psi \{\lambda_a\}\rangle$ of the single particle Hamiltonian $H_{\rm MF}$ can be gotten by diagonalization. The variational parameters are determined by minimizing the total energy
\begin{equation}
    E_{\rm T} = \langle \psi \{\lambda_a \} | \hat{H}_{\rm T} | \psi \{\lambda_a\} \rangle.
\end{equation}
The form of $\hat{H}_{\rm T}$ is written in the main text. To start the minimization procedure, one must set initial values of the variational parameters $\{\lambda_a\}$. Notice that the minimization results will definitely depend on the initial input. In our work, for each interaction parameter, we perform several calculations starting from all kinds of ordered inputs, including QAH, NMI, charge stripe orders and so on. Besides, inputs with random values have also been performed. The ground state is determined by comparing the total energy starting from different input values.

The results have been checked by the self-consistent mean filed program, which gives out the same results. In the self-consistent procedure, the four-fermion operators in $\hat{H}_{\rm T}$ are decoupled as
\begin{equation}
\label{mean}
\begin{split}
        \hat{c}^{\dagger}_i \hat{c}^{\dagger}_j \hat{c}_k \hat{c}_l \mapsto 
             &\langle \hat{c}^{\dagger}_j \hat{c}_k \rangle \hat{c}^\dagger_i \hat{c}_l +
              \langle \hat{c}^{\dagger}_i \hat{c}_l \rangle \hat{c}^\dagger_j \hat{c}_k -
              \langle \hat{c}^{\dagger}_i \hat{c}_l \rangle\langle\hat{c}^\dagger_j \hat{c}_k \rangle \\
             &-\langle \hat{c}^{\dagger}_j \hat{c}_l \rangle \hat{c}^\dagger_i \hat{c}_k -
              \langle \hat{c}^{\dagger}_i \hat{c}_k \rangle \hat{c}^\dagger_j \hat{c}_l +
              \langle \hat{c}^{\dagger}_i \hat{c}_k \rangle\langle\hat{c}^\dagger_j \hat{c}_l \rangle.
\end{split}
\end{equation}
The expectation values in above equation are calculated self-consistently after initial values are given. Besides, the self-consistent mean field program has been used to get the effective single-particle Hamiltonian and wave function under the mean field approximation. Different from the variational method, the self-consistent procedure keeps the proper symmetry of the Hamiltonian and wave function. The edge state and the parity of the occupied states are calculated by using the self-consistent mean field results.

%\section{Hartree-Fock results with only V}

\begin{figure}[!htb]
	\includegraphics[width=0.95\textwidth]{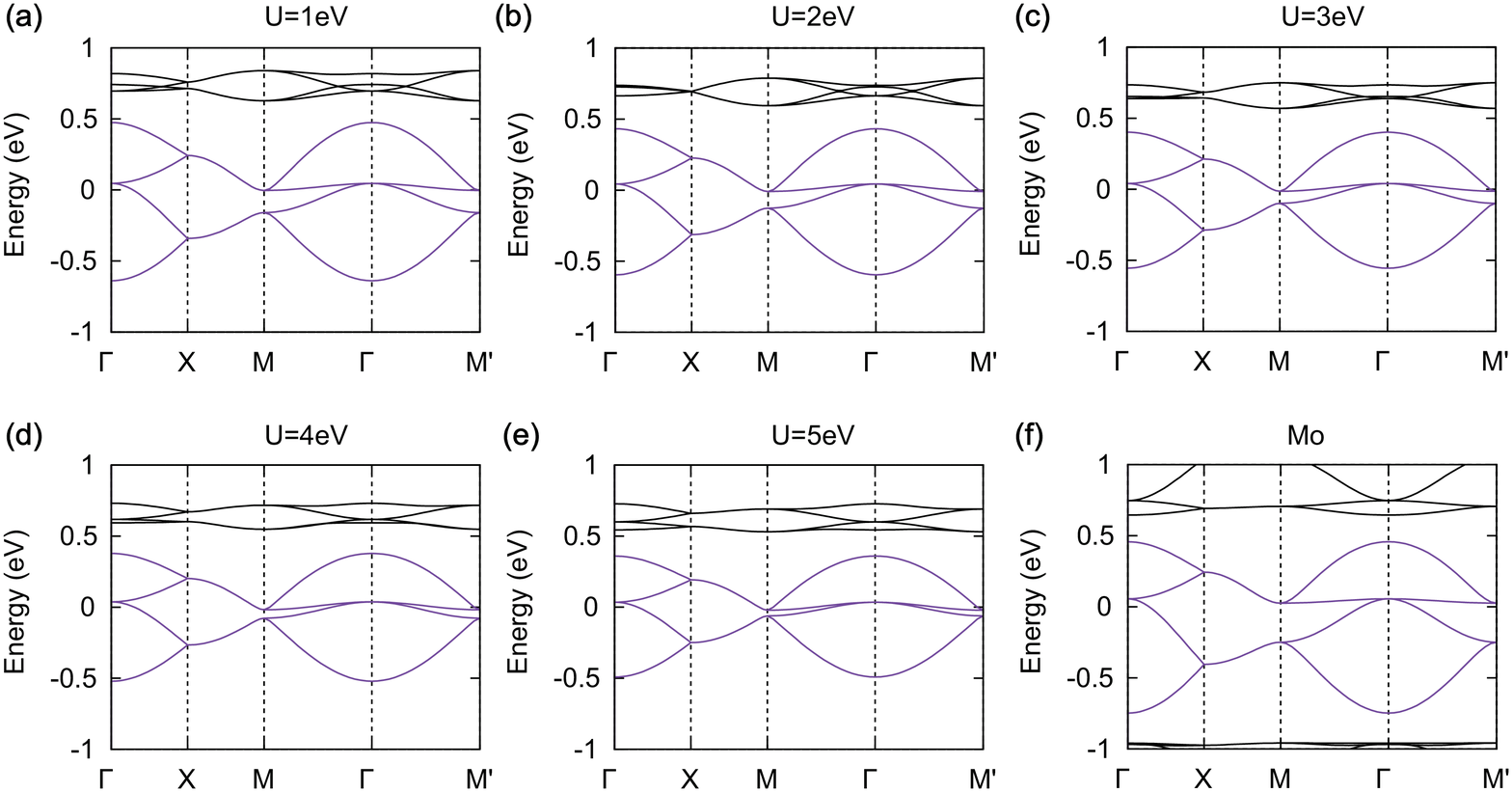}
	\caption{ The DFT band structure of ferromagnetic \Cr with different value of $U$. The black and purple lines represent the spin up and spin down polarized bands, respectively. (a)-(e) The band width of the spin-down bands around Fermi level is slightly decreased with the increasing value of $U$. (f) The DFT band structure of ferromagnetic \Mo.
	}
	\label{fig:UMo}
\end{figure}

\end{widetext}

\end{document}